\documentstyle[preprint,prd,aps,tighten,epsf]{revtex}
\begin{document}
\draft
\newcommand{\nl}{\nonumber \\}
\newcommand{\bea}{\begin{eqnarray}}
\newcommand{\eea}{\end{eqnarray}}
\newcommand{\barr}{\begin{array}}
\newcommand{\earr}{\end{array}}
\newcommand{\bi}{\bibitem}
\newcommand{\be}{\begin{equation}}
\newcommand{\ee}{\end{equation}}
\def\ra{\rightarrow}
\def\dek#1{\times10^{#1}}
\def\kdec{K^0\ra K^+ e^- \bar\nu_e} 
\def\akdec{\bar{K}^0\ra K^- e^+ \nu_e} 
\def\kldec{K^0_L\ra K^\pm e^\mp \nu}
\def\ksdec{K^0_S\ra K^\pm e^\mp \nu}
\def \prd#1#2#3{Phys.~Rev.~D~{\bf#1}, #2 (#3)}
\def \prl#1#2#3{Phys.~Rev.~Lett.~{\bf#1}, #2 (#3)}
\def \ea{{\it et al.}}
\def \epjc#1#2#3{Eur.~Phys.~J.~C~{\bf#1}, #2 (#3)}
\def \plb#1#2#3{Phys. Lett. B~{\bf#1}, #2 (#3)}

\title{
{\bf 
Can the beta decay of neutral kaons be observed?\\
}}
\author{Peter Lichard$^{1,2}$
and Julia Thompson$^1$}
\address{
$^1$Department of Physics and Astronomy, University of Pittsburgh,
Pittsburgh, Pennsylvania 15260 \\
$^2$Institute of Physics, Silesian University, 746-01 Opava, Czech Republic
}
\maketitle
\begin{abstract}
The rate of the beta decay of neutral kaons is calculated within
the meson dominance approach taking into account the relation between the 
$K K\rho$ and $\pi\pi\rho$ coupling constants which follows from the 
vector meson dominance in electromagnetic interactions and isospin symmetry. 
The decay rate transforms into the following branching fraction summed over 
the charge states indicated: $B(K^0_L\ra K^\pm e^\mp\nu)=(2.53\pm0.10)
\times10^{-9}$. The error is dominated by our estimate of isospin violating
effects. Experimental aspects of such a measurement are discussed.
\end{abstract}

\narrowtext
In this work we built upon previous work by one of us \cite{md} to give
a prediction for the $\kldec$ branching fraction which may be within the 
reach of some next generation neutral rare kaon decay experiments. The 
calculation presented here is performed using the meson dominance (MD) 
in  weak interactions and the vector meson dominance (VMD) in electromagnetic 
interactions. Both these hypotheses naturally stem \cite{md} from the 
Standard Model of electroweak interactions. The measurement of the
$\kldec$ decay can thus be considered as another test of the Standard
Model.

The MD hypothesis leads \cite{md} to the following 
formula for the $\kdec$ and $\akdec$ differential decay rate (see Fig.~1 
for the corresponding Feynman diagram) 
\be
\label{drate}
\frac{d\Gamma}{dt}=\frac{G_F^2X_{K^0K^+\rho^-}}
{3(4\pi m_{K^0})^3}\ \frac{t-m_e^2}{t^3}\ \sqrt{\lambda(t)}
\left[\varphi_1(t)-\varphi_2(t)\right]\left(\frac{r}{r-t}\right)^2\ ,
\ee
where $t$ is the invariant mass squared of the outgoing lepton pair,
$r$ is the mass of the $\rho$ meson squared, and
\be
\label{x}
X_{K^0K^+\rho^-}=\left|V_{ud}\right|^2\left(\frac{g_{K^0K^+\rho^-}}
{g_\rho}\right)^2\ ,
\ee
with $g$s denoting the strong coupling constants. For the $g_{\pi\pi\rho}$
coupling constant, a simpler notation $g_\rho$ is used. Furthermore,
\bea
\label{lambda}
\lambda(t)&=&t^2+x^2+y^2-2tx-2ty-2xy\ ,\\
\label{phi1}
\varphi_1(t)&=&2t^4-(4x+4y+z)t^3+\left[2(x-y)^2+z(2x+2y-z)
\right]t^2\nl
&&+2z\left[(x-y)^2+z(x+y)\right]t  - 4z^2(x-y)^2\ ,\\ 
\label{phi2}
\varphi_2(t)&=&\frac{3tz}{r^2}(2r-t)(t-z)(x-y)^2 .
\eea
Here, $x=m_{K^0}^2$, $y=m_{K^+}^2$, and $z=m_e^2$. The quantity (\ref{x}) was 
estimated in Ref.~\cite{md} using the experimental
rate of the $\tau^-\ra K^-K^0\nu_\tau$ decay. The value obtained,
$X_{K^0K^+\rho^-}=0.64\pm0.12$,
led to the following branching fraction prediction 
$B(\kldec)=(3.4\pm0.6)\dek{-9}$,
where the sum over the charge states indicated is understood. 

In  more recent work \cite{kpiee_kp} one of us  has shown that the VMD 
hypothesis, together with 
the isospin invariance of the strong vertices and normalization conditions 
on the form factors, implies that the contribution of the $\rho$ meson pole 
to the $K^+$ form factor is just half of that to the $\pi^+$ form factor. 
In terms of the coupling constants it means that
\be
\label{ccratio}
\frac{g_{K^+K^+\rho^0}}{g_\rho}=\frac{1}{2}\ .
\ee
To establish the connection of this result with our quantity (\ref{x}) 
the isospin invariance relation $|g_{K^+K^0\rho^+}|=\sqrt{2}\ |g_{K^0K^+
\rho^-}|$ is required. We thus obtain
\be
\label{xnew}
X_{K^0K^+\rho^-}=\frac{1}{2}\left|V_{ud}\right|^2=0.4739\pm0.0008,
\ee
where the numerical value of $|V_{ud}|$ comes from Ref.~\cite{pdg}. 

Since the relative errors of all parameters that serve as input 
to our formula (\ref{drate}) are very small, it is important to discuss
the systematic errors of our approach. They consist of several components.

First, we should account somehow for a possible violation of the isospin 
symmetry, which was assumed to be unbroken when deriving Eq.~(\ref{xnew}). 
Of course, there is no exact way to do that. Looking at the various
quantities which reflect the violation of isospin invariance {\bf (}$n-p$, 
$\pi^+-\pi^0$, and $K^0-K^+$  mass differences, $K^{*+}-K^{*0}$
width difference, violation of the $SU(2)$ relation between the $K_{e3}^+$
and $K_{e3}^0$ decay rates{\bf )} we expect that the relative error connected 
with this 
phenomenon does not exceed four per cent, which we will take as our very
conservative ``educated guess". 

Second, the systematic error of the MD approach is almost certainly
negligible in this case. We base our judgement on a very precise MD result 
\cite{md} for the branching fraction of the $\pi^+\ra\pi^0e^+\nu_e$ decay, 
which is a process analogous to that considered here. In both cases we
are very close to the threshold, where the assumptions leading to
the MD formulas are well satisfied \cite{md}.  

Third, there are uncertainties in the form of the $\rho$ meson propagator. 
One of us discussed this issue in Refs.~\cite{ratio,k0star}, where references 
to some previous papers can also be found. Here, we proceed in the same way 
as in ~\cite{k0star}. We  calculate the integrated branching fraction twice, 
for two  
(extreme) choices of the $\rho$ meson propagator and consider the difference
between the two results as a measure of the systematic error. One of the 
choices is equivalent to using our original formulas (\ref{drate}-\ref{phi2}) 
and the other to putting $\varphi_2(t)$, Eq. (\ref{phi2}),  to zero 
identically. Both choices lead to the same result\footnote{This can easily
be understood because $\varphi_2(t)$ is proportional to $(m_e/m_\rho)^2$.}, 
which after the conversion to units of inverse seconds reads:
\be
\label{gamma}
\Gamma(\kdec)\equiv\Gamma(\akdec)=
(4.887\pm0.008)\dek{-2}\ {\rm s}^{-1}.
\ee
The error quoted here reflects only the error of the $V_{ud}$ element
of the Cabibbo-Kobayashi-Maskawa matrix. So, the systematic error connected
with the uncertainties in the $\rho$ meson propagator is also negligible.

Using Eq. (\ref{gamma}) and the experimental value of the $K^0_L$ lifetime 
\cite{pdg} we arrive at the branching fraction
\be
B(\kldec)=(2.53\pm0.10)\dek{-9}.
\ee
The error is dominated by our ``educated guess" of isospin
violating effects. 

For the other pseudoscalar mesons, beta decays  (i.e. semileptonic transitions between
the members of the same isotopic multiplets) 
may completely escape detection  because of their extremely small branching
fractions. The low-lying cases are discussed briefly below.

The branching fraction of the $K_S^0$ state, $B(\ksdec)=(4.37\pm0.17)
\dek{-12}$, comes  from Eq.~(\ref{gamma}) and the experimental value of 
the $K_S^0$ mean lifetime. The error is based on our guess above.

When calculating the decay rate of the $D^+\ra D^0e^+\nu_e$ decay, we can 
use the same formalism as we have applied to the beta decay of neutral kaons.
The electromagnetic form factors of the $D^+$ and $D^0$ 
mesons now contain, in addition to the $\rho$, $\omega$, and $\phi$ terms, 
an important contribution from the $J/\psi$ pole. But the $J/\psi$ term
transforms like an isoscalar and therefore again only the $\rho$ meson pole
contributes to the isovector part of the $D^+$ and $D^0$ form factors.
And it is this condition, not the detailed structure of the isoscalar part,  
which leads, together with the normalization of the $D^+$ and $D^0$ 
electromagnetic 
form factors, to the relation between the strong coupling constants of the 
same kind as shown in Eq. (\ref{ccratio}). The value of $X_{D^+D^0\rho^+}$ 
is thus the same as that given in Eq.~(\ref{xnew}). The decay rate 
comes out about five times larger than that shown in Eq.~(\ref{gamma}), 
as a result of the larger mass difference and increased phase space. 
But the branching fraction is pushed to an unobservable value, 
$B(D^+\ra D^0e^+\nu_e)\approx 2.7\dek{-13}$, by the very short 
$D^+$ lifetime. 

In the $(B^+,B^0)$ isomultiplet, the mass difference is smaller
than the electron mass, so the beta transition is not possible.

Above we have discussed theoretical expectations for the branching fraction 
of the $K^0_L$ beta decay. Next we will briefly summarize some signatures of 
this decay and those accelerators and experiments which might have a chance 
to observe it.

First we note that the small $K^0-K^+$ difference ($\delta m\approx 4$~MeV) 
gives striking characteristics\footnote{The formulas used below are valid if
the beam momentum is much greater than $\delta m$, which is satisfied
in the cases we consider.} to the kaon beta decay events:
\begin{enumerate}
\item
a charged kaon with momentum approximately that of the beam
(the relative loss of momentum is less than $\delta m/m_{K^0}$)
and very close to the initial direction of the beam 
($\theta < \delta m/p_{K^0}$, where $p_{K^0}$ is the beam
momentum).

\item
a large-angle electron (or positron) with small momentum
($p<(E_{K^0}+p_{K^0})\delta m/m_{K^0}$, where $E_{K^0}$ is
the beam energy).

\end{enumerate}

Although these characteristics are striking, they are not simple
to achieve. Many experiments are intentionally blind in the forward
direction, where the kaon will be found.  Not all experiments are in 
a position to measure the momentum of the incident kaon.  And not 
all experiments have a magnetic spectrometer or other means to measure
the momentum and angle of the charged decay products. Even for experiments 
with magnetic spectrometers and information on the incident kaon momentum
and direction, the suppression of the $\pi^\pm e^\mp\nu$ modes, which are
about $10^8$ times more frequent than $K^\pm e^\mp\nu$, requires
particle identification to discriminate between pions and kaons. 

Because of the intense kaon beam requirement, there are 
only a few facilities in the world at which the kaon beta decay 
could be measured: Brookhaven National Laboratory (BNL), Fermi National
Accelerator Laboratory (Fermilab), European Organization
for Nuclear Research (CERN), Phi-factory DA$\Phi$NE at the Frascati National
Laboratory, and the High Energy Accelerator Research Organization (KEK) in 
Tsukuba.  

A natural place for the kaon beta decay measurement would be a beam 
designed for the important decay $K^0_L\ra\pi^0\nu\bar\nu$ and similar rare
decays of neutral kaons. Such beams are presently in place or in the 
planning or preparatory stages at all the laboratories mentioned above. 
The characteristics and advantages of the kaon beams and detectors 
vary:

(1) The completed experiment 
E871 at BNL was a search for the lepton family violating decay 
$K^0_L\ra\mu^\pm e^\mp$, with a limit set at the $4.7\dek{-12}$ level
\cite{k0lmue} and observation of the  $K_L\ra e^+e^-$ branching fraction as
$(8.7^{+5.6}_{-4.1})\dek{-12}$ \cite{k0lee}.  This experiment had a double 
charged particle spectrometer with electron and muon identification. However,
it was blind in the forward direction, and the kaon beta decay cannot 
be studied in their data \cite{Wojcicki}.

(2) The KTeV experiment at Fermilab, \cite{ktev}) has a  beam of 70~GeV/c
neutral kaon average momentum, leading to charged kaons of approximately 
70~GeV/c momentum and nearly exactly straight forward (within about 0.06~mrad)
and electrons up to about 1.1~GeV/c (the estimates here and below are given 
for the average beam momentum). The momentum kick of the spectrometer 
magnet (205~MeV/c) would allow the highest energy electrons to pass through, 
but would bend the kaons only by about 3~mrad, not enough to avoid the
50~mrad blind spot in the center of the calorimeter. The  kaons
might still be seen by the drift chambers, and some $\pi/K$
rejection might be available from the transition radiation detectors, 
though they are optimized
for $e/\pi$ separation at lower energies.  But these events would not
be included in any data taken to date because of the trigger requirement
of $\approx20$~GeV/c in the calorimeter. There is no time of flight and
therefore no momentum information for the parent kaon.

(3) Experiment NA48 at CERN \cite{na48} has a kaon beam of 110 GeV/c,
which, as for KTeV, would lead to charged kaons nearly straight ahead 
($\theta < 0.04$~mrad). The latter thus remain in the beam pipe and
cannot reach the detector. The electron momentum extends up to 1.8~GeV/c.
Experiment NA48 has a magnetic spectrometer but no time of flight for the 
incident kaon, nor particle identification to distinguish charged kaons 
and pions.

(4) The KLOE detector at DA$\Phi$NE \cite{daphne} is a general purpose 
detector with $K_{S} K_{L}$ pairs produced nearly at rest by decays from 
the $\phi$(1020).
The momentum is known, and the opposite kaon is measured, to give the
kaon direction.  The decay kaon may be at angles of order 36~mrad,
but electron momenta will be only of order 5~MeV/c,  below the
spectrometer minimum momentum for detection, in normal running.

(5) For the KOPIO experiment \cite{kopio} BNL has proposed an intense 
neutral kaon beam of 0.7 GeV/c average momentum, 
with time-of-flight tagging allowing momentum determination of the $K^0$
to within  a few  per cent.  The advantage of such momentum
resolution is important for the $K^0_L$ beta decay.  The outgoing kaon 
would have an angle ranging out to about 6~mrad from the incident beam,
and the electron momentum would range up to about 13~MeV/c. The KOPIO
experiment as proposed does not include any provision for a magnetic
spectrometer (or any other charged particle identification except for 
identification of electrons and positrons in the electromagnetic 
calorimeter).

(6) Fermilab has a proposed experiment KAMI (Kaons at the Main Injector) 
\cite{kami} with $K^0$ momentum about 10~GeV/c. The momentum of this beam 
is too high for time of flight  momentum measurement of the beam, but the 
proposed experiment has a charged particle spectrometer which could detect 
a fast outgoing charged particle close to the beam and a slow wide angle 
particle appropriate for the electron or positron.
No charged particle identification is included at this time, beyond the
comparison of energy deposit with momentum measured in the spectrometer.
The kaon would be within about 0.4 mrad of the incident $K^0$ direction,
and the electron momentum would range up to about 160 MeV/c.
This experiment is essentially an upgraded continuation of the KTeV
experiment, with nearly the same collaborators.

(7) In KEK, a $K^0_L\ra\pi^0\nu\bar\nu$ experiment E391A is in preparation
\cite{kek} by experimenters from KEK, Osaka University, Saga University, 
and Yamagata University. It will be situated in a new $K^0$ beamline at the 
12~GeV Proton Synchrotron. The detection system consists of an array of
CeF$_3$ crystal calorimeters to measure the energies and positions of the 
two gammas from the $\pi^0$ decay, and a barrel of lead-scintillator sandwich 
calorimeters to eliminate backgrounds involving other particles. The average
kaon momentum will be about 2~GeV/c. The momentum of the electron from
the kaon beta decay would range up to about~33 MeV/c, and the outgoing 
charged kaon would be within about 2~mrad of the incident $K^0$ direction.
Since neither time of flight for the kaon momentum nor charged particle 
identification or measurement is planned, it seems unlikely that the KEK 
experiment as designed will observe the kaon beta decay. Let us note
that a similar experiment is envisioned \cite{jhfkpinunu} at the planned JHF 
(Japan Hadron 
Facility) 50 GeV high intensity proton synchrotron,  which is a joint 
project of JAERI (Japan Atomic Energy Research Institute) and KEK \cite{jhf}.  

In summary, the (as yet unmeasured) kaon beta decay  has a predicted 
branching fraction of $(2.53\pm0.10)\dek{-9}$ within the MD model.
Observation of the decay at this level would be a test of 
meson dominance.  A substantial departure from this prediction
would be surprising within the framework of the standard model.
If the elusive $K^0_L\ra\pi^0\nu\bar\nu$  mode is measured,
and if some thought is given beforehand to details of the experimental
configuration, the measurement of the kaon beta decay branching fraction 
may be a useful and interesting additional result.  
The two experiments with the most intense neutral kaon beams
are KAMI at Fermilab and KOPIO at BNL.  The background from
the semileptonic decay  to $\pi e \nu$ is of order $10^{8}$ times
the expected signal.  No planned experiment has all three possible
handles: incident kaon momentum and angle information; outgoing
charged decay particle spectrometer; and particle identification to
identify kaons and pions in the charged decay particles.
Detailed studies beyond the scope of this note would be  required
to establish whether either the KAMI 
experiment (which has a charged particle spectrometer but no decay  particle
identification and no time of flight for the incident momentum) or
the KOPIO experiment (which has time of flight for the incident kaon momentum
but no charged particle spectrometer and no charged decay product
particle identification) could be successful without modification of their
apparatus.

\acknowledgements
We are indebted to Elliott Cheu, Bob Hsiung, Konrad Kleinknecht, Dave Kraus, 
Ivan Mikulec, and Mike Zeller for useful discussions. This work was supported 
by the U.S. Department of Energy under contract No. DOE/DE-FG02-91ER-40646 
and by the Grant Agency of the Czech Republic under contract No. 202/98/0095.

\begin{figure}
\begin{center}
\leavevmode
\setlength \epsfxsize{15cm}
\epsffile{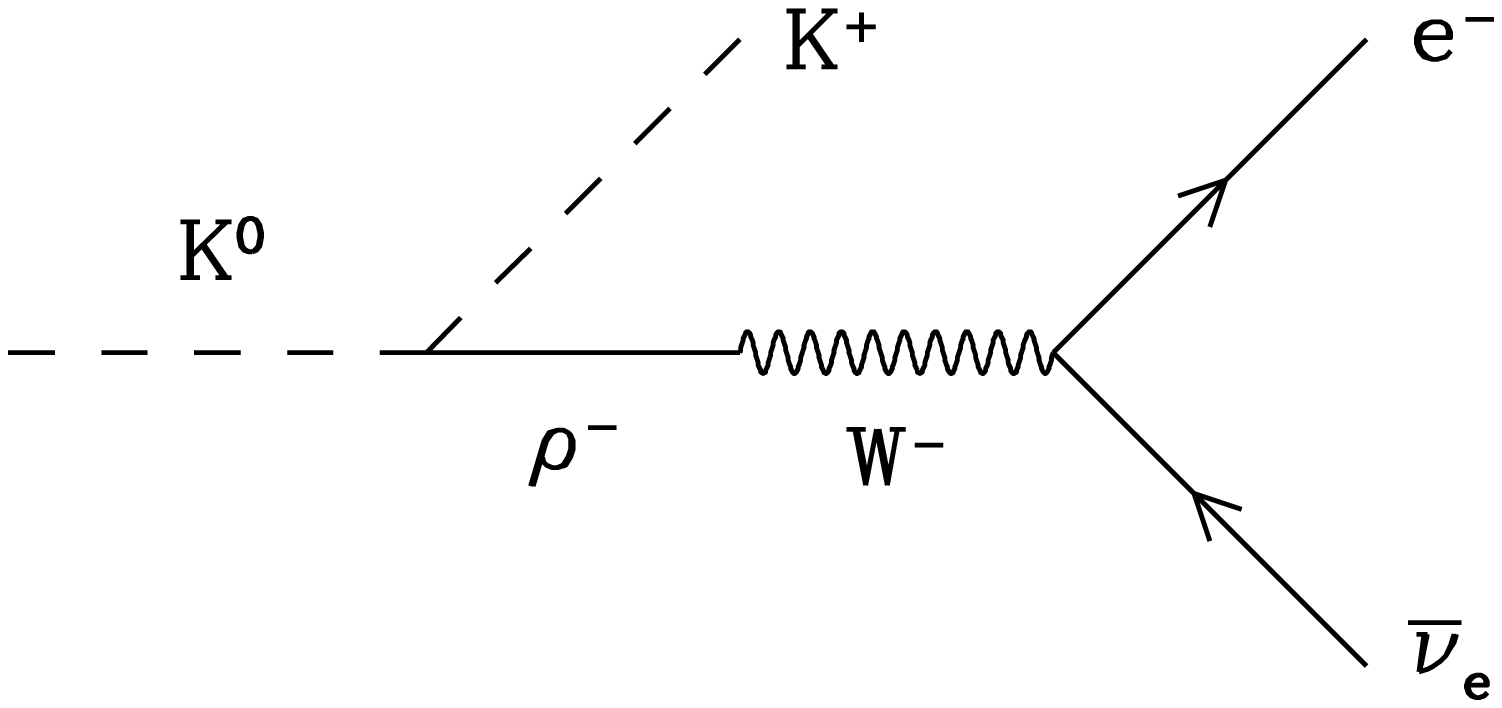}
\end{center}
\caption{Feynman diagram of the $K^0\ra K^+ e^-\bar\nu_e$
decay in MD approach.}
\label{kkenu}
\end{figure}

\end{document}